\documentclass{webofc}
\usepackage[varg]{txfonts}   
\pdfoutput=1
\usepackage{pstricks}
\usepackage{color}
\usepackage{amssymb,amsmath,bbm}
\usepackage{epsf}
\usepackage{epsfig}
\usepackage{afterpage}
\usepackage{longtable}
\usepackage{latexsym,mathrsfs,dsfont}
\usepackage{graphics}
\usepackage{url}
\usepackage{paralist}
\usepackage{bbold}

\newcommand{\tev}{\, {\rm TeV}}
\newcommand{\gev}{\, {\rm GeV}}

\newcommand{\ms}{m_{\rm s}}
\newcommand{\md}{m_{\rm d}}

\newcommand{\vcb}{|V_{cb}|}

\newcommand{\vub}{|V_{ub}|}

\newcommand{\ord}{\mathcal{O}}
\newcommand{\bsi}{B_6^{(1/2)}}
\newcommand{\bei}{B_8^{(3/2)}}

\def\epe{\varepsilon'/\varepsilon}
\newcommand{\be}{\begin{equation}}
\newcommand{\ee}{\end{equation}}
\newcommand{\bi}{\begin{itemize}}
\newcommand{\ei}{\end{itemize}}

\usepackage{tikz}
\newcommand*\circled[1]{\tikz[baseline=(char.base)]{
            \node[shape=circle,draw,inner sep=2pt] (char) {#1};}}

\woctitle{QCD@Work 2014}
\begin{document}

\title{\boldmath $\Delta I=1/2$ Rule and $\hat B_K$ : 2014 \unboldmath}

%
%

\author{ Andrzej~J.~Buras\inst{1,2}\fnsep\thanks{\email{aburas@ph.tum.de}} 
}

\institute{TUM Institute for Advanced Study, Lichtenbergstr. 2a, D-85748 Garching, Germany
\and
           Physik Department, Technische Universit\"at M\"unchen,
James-Franck-Stra{\ss}e, D-85748 Garching, Germany
          }

\abstract{%
  I summarize the status of the $\Delta I=1/2$ rule in  $K\to\pi\pi$ decays  
within an {\it analytic}  approach based on the dual representation of QCD as a theory of weakly interacting mesons for large $N$, where $N$ is the number of colours. This approximate approach, developed in the 1980s by 
William Bardeen, Jean-Marc G\'erard and myself, allowed us  already 28 years ago to identify
the dominant dynamics  behind the $\Delta I=1/2$ rule. However,  
the recent inclusion of lowest-lying vector meson contributions in addition to the pseudoscalar ones to hadronic matrix elements of current-current operators and the calculation of the corresponding Wilson coefficients in a momentum scheme at the NLO 
improved significantly the matching between quark-gluon short distance contributions and meson long distance contributions over our results in 1986.  We obtain satisfactory 
description of the ${\rm Re}A_2$ amplitude and 
${\rm Re}A_0/{\rm Re}A_2=16.0\pm 1.5$ to be compared  with its experimental value of $22.3$. 
 While this difference could be the result of present theoretical uncertainties in  our approach, it cannot be excluded that 
 New Physics (NP) is here at work. The analysis by
Fulvia De Fazio, Jennifer Girrbach-Noe and myself shows that indeed 
a tree-level 
$Z^\prime$ or $G^\prime$ exchanges with  masses in the reach of the LHC and special couplings 
to quarks can significantly improve the theoretical status of the $\Delta I=1/2$ rule while satisfying constraints from $\varepsilon_K$, $\epe$, $\Delta M_K$, LEP-II and the LHC. The ratio $\epe$ plays an important role in these 
considerations. I stress that our approach allows to understand the physics 
behind recent numerical results obtained in lattice QCD not only for the $\Delta I=1/2$ rule but also for the parameter $\hat B_K$ that enters the evaluation of $\varepsilon_K$. In contrast to the $\Delta I=1/2$ rule and $\epe$ the chapter 
on $\hat B_K$ in QCD appears to be basically closed.
} 

\maketitle
\section{Introduction}
\label{intro}
One of the puzzles of the 1950s was a large disparity between the measured 
values of the real parts of the isospin amplitudes $A_0$ and $A_2$ 
for a kaon to decay into two pions which on the basis of usual isospin 
considerations were expected to be of the same order. In 2014 we know the 
experimental values of the real parts of these amplitudes very precisely 
\cite{Beringer:1900zz}
\be\label{N1}
{\rm Re}A_0= 27.04(1)\times 10^{-8}~\gev, 
\quad {\rm Re}A_2= 1.210(2)   \times 10^{-8}~\gev.
\ee
As ${\rm Re}A_2$ is dominated by $\Delta I=3/2$ transitions but 
${\rm Re}A_0$ receives contributions also from $\Delta I=1/2$ transitions, 
the latter transitions dominate ${\rm Re}A_0$  which expresses
the so-called $\Delta I=1/2$ rule \cite{GellMann:1955jx,GellMann:1957wh}
\be\label{N1a}
R=\frac{{\rm Re}A_0}{{\rm Re}A_2}=22.35.
\ee
In the 1950s QCD and  Operator Product Expansion did not exist and clearly 
one did not know that $W^\pm$ bosons existed in nature but using the ideas 
of Fermi, Gell-Mann, Feynman, Marshak and Sudarshan one could still 
evaluate the amplitudes ${\rm Re}A_0$ and ${\rm Re}A_2$ to find out 
that such a high value of $R$ is a real puzzle.

In modern times we can reconstruct this puzzle by evaluating the 
simple $W^\pm$ boson exchange between the relevant quarks which after integrating out $W^\pm$ 
generates the current-current operator $Q_2$:
\begin{equation}\label{O1} 
Q_1 = (\bar s_{\alpha} u_{\beta})_{V-A}\;(\bar u_{\beta} d_{\alpha})_{V-A},
\qquad Q_2 = (\bar su)_{V-A}\;(\bar ud)_{V-A}~.
\end{equation}
We have listed here the second current-current operator, $Q_1$, which we will 
need soon. With only $Q_2$ contributing we have 
\be
 {\rm Re}A_{0,2}=\frac{G_F}{\sqrt{2}}V_{ud}V_{us}^*\langle Q_2\rangle_{0,2}\,.
\ee
Calculating  the matrix elements $\langle Q_2\rangle_{0,2}$ in the strict large $N$ limit, which corresponds to factorization of matrix elements of $Q_2$ into 
the product of matrix elements of currents, we find 
\be\label{LO}
{\rm Re}A_0=3.59\times 10^{-8}\gev ,\qquad   {\rm Re}A_2= 2.54\times 10^{-8}\gev~, \qquad R=\sqrt{2}
\ee
in plain disagreement with the data in (\ref{N1}) and (\ref{N1a}). 
It should be emphasized that the explanation of the  missing enhancement factor of $15.8$ in $R$ through some dynamics must simultaneously give the correct values for ${\rm Re}A_0$ and  ${\rm Re}A_2$. 
This means that this dynamics should suppress  ${\rm Re}A_2$ by a factor of $2.1$, not more, and enhance ${\rm Re}A_0$ by a factor of $7.5$. 

It is evident that what is missing in this calculation are strong interaction 
effects represented these days by QCD but the question arises whether the 
physical picture behind the $\Delta I=1/2$ rule as described by QCD has 
a simple structure. As demonstrated by Bardeen, G\'erard and myself already 
in 1986 \cite{Bardeen:1986vz} and improved on the technical level by us recently 
\cite{Buras:2014maa} the dominant dynamics behind the $\Delta I=1/2$ rule 
has in fact a simple structure.

To this end one should note that from the point of view of operator product 
expansion the calculation we have just performed to get (\ref{LO}) corresponds to 
\begin{itemize}
\item
The evaluation of the Wilson coefficient of the operator $Q_2$ in a free (from the point of view of strong interactions) theory of quarks, which corresponds to scales $\mu=\ord(M_W)$ and setting $\alpha_s(M_W)=0$. 
\item
The evaluation of hadronic matrix elements $\langle Q_2\rangle_{0,2}$ in a
free theory of mesons which corresponds to the factorization scale $\mu=\ord(m_\pi)\approx 0$ and setting $N$ to infinity.
\end{itemize}

The second point follows from 
the dual representation of QCD as a theory of weakly interacting mesons for large $N$, advocated already in the 1970s in \cite{'tHooft:1973jz,'tHooft:1974hx,Witten:1979kh,Treiman:1986ep}. In the strict large $N$ limit QCD becomes a free theory 
of 
mesons and in this limit the calculation of hadronic matrix elements by 
means of factorization method is correct within QCD \cite{Buras:1985xv}.
 But as the Wilson 
coefficient of $Q_2$ has been evaluated at $\mu=\ord(M_W)$ and its  hadronic matrix elements at  $\mu=\ord(m_\pi)\approx 0$ our calculation of  ${\rm Re}A_0$ and  ${\rm Re}A_2$ is incomplete. In order to complete it we have to fill
the gap between these two 
vastly different energy scales with QCD dynamics represented by 
quark-gluon interactions at short distance scales and by meson interactions 
at long distance scales. This requires the inclusion of $\alpha_s$ effects 
at short distances and $1/N$ corrections in the meson theory at long 
distances.

In Section~\ref{sec:2} I will describe the structure of our approach together 
with results for the $A_{0,2}$ amplitudes in three steps and will compare it with the lattice QCD approach. In this context I will also summarize the status of 
the parameter $\hat B_K$.
In Section~\ref{sec:3} I will summarize an analysis performed by Fulvia 
De Fazio, Jennifer Girrbach-Noe and myself which demonstrates that
tree-level $Z^\prime$ or $G^\prime$ exchanges with  masses in the reach of the LHC and special couplings 
to quarks can significantly improve the theoretical status of the $\Delta I=1/2$ rule while satisfying constraints from $\varepsilon_K$, $\epe$, $\Delta M_K$, LEP-II and the LHC. Few comments in Section~\ref{sec:4} close this brief review.
I am presenting here the way I see the dynamics behind the $\Delta I=1/2$ rule. 
Over the years other views have been expressed in the literature. See in particular \cite{Stech:1990iu} and most recent papers \cite{Crewther:2013vea,Liu:2014vha} where further references can be found.

\boldmath
\section{The Dynamics behind the $\Delta I=1/2$ Rule}\label{sec:2}
\unboldmath
\subsection{Step 1: Quark-Gluon Evolution}
This step involves the calculation of the Wilson coefficients $z_{1,2}$ of the 
current-current operators $Q_{1,2}$  at a low energy scale $\mu=\ord(1\gev)$ and fills the gap present in our simple calculation between this scale and 
the electroweak scale $\ord(M_W)$.  Having them one can calculate 
 the $K\to\pi\pi$ decay
amplitudes in the Standard Model  using
\be\label{basic}
A(K\to\pi\pi)=\frac{G_F}{\sqrt{2}}V_{ud}V_{us}^*\sum_{i=1,2} z_i(\mu)\langle \pi\pi|Q_i(\mu)|K\rangle,
\ee
where QCD penguin contributions have been omitted as they will be included 
in Step 3 below. We have indicated that the matrix elements are to be evaluated 
at $\mu=\ord(1\gev)$ but in this step we will still keep them at $\mu\approx 0$ 
and use their values calculated in the strict large $N$ limit. We will improve 
on this in Step 2.

The coefficients $z_i(\mu)$ have been calculated at leading order in the 
renormalization group improved perturbation theory in \cite{Gaillard:1974nj,Altarelli:1974exa}. This pioneering calculations of short distance QCD effects have shown that these effects indeed enhance ${\rm Re}A_0$ and suppress ${\rm Re}A_2$. However, the inclusion of NLO QCD corrections to $z_{1,2}$ \cite{Altarelli:1980fi,Buras:1989xd}
made it clear, as stressed in particular in \cite{Buras:1989xd}, that the 
 $K\to\pi\pi$ amplitudes without the proper calculation of hadronic matrix 
elements of $Q_i$ are both scale and renormalization scheme dependent. For instance setting $\mu=0.8\gev$ we find
\be
R_{cc}({\rm NDR-\overline{MS}})\approx 3.0,\qquad R_{cc}({\rm \overline{MOM}})\approx 4.4 \,,
\ee
where the subscript {\it cc} indicates that only current-current contributions 
have been taken into account and  ${\rm \overline{MOM}}$ is a momentum 
scheme, introduced in \cite{Buras:2014maa}, which is particularly suited for 
the calculations of the amplitudes in our approach. In this scheme one finds 
then for $\mu=0.8\gev$
\be\label{NLO}
{\rm Re}A_0=7.1\times 10^{-8}\gev ,\qquad   {\rm Re}A_2= 1.6\times 10^{-8}\gev~. 
\ee
This is a significant improvement over the results in (\ref{LO}) bringing the 
theory closer to the data in (\ref{N1}) and (\ref{N1a}). However, this result 
is scale and renormalization scheme dependent. For ${\rm NDR-\overline{MS}}$ 
scheme and $\mu\approx (2-3)\gev$ as used in lattice QCD calculations this 
improvement would be much smaller. But, even in  ${\rm \overline{MOM}}$ scheme 
and at $\mu=0.8\gev$,  further enhancement 
of ${\rm Re}A_0$ and further suppression of ${\rm Re}A_2$ are needed in order 
to be able to understand the $\Delta I=1/2$ rule. This brings us to Step 2 
which fills the remaining gap in our original calculation.

\subsection{Step 2: Meson Evolution}
 The renormalization group evolution down to the scales $\ord(1\gev)$ just 
performed is continued as a short but fast meson evolution down to zero momentum scales at which the factorization of hadronic matrix elements is at work. 
Equivalently, starting with factorizable hadronic matrix elements 
$\langle Q_1 \rangle_{0,2}$ and $\langle Q_2 \rangle_{0,2}$ at $\mu\approx 0$ 
and evolving them to $\mu=\ord(1\gev)$ at which $z_{1,2}$ are calculated one is able 
to calculate the matrix elements of these two operators at $\mu=\ord(1\gev)$ and properly 
combine them with  $z_{1,2}$ calculated in the ${\rm \overline{MOM}}$ scheme.
Details of these calculations can be found in \cite{Bardeen:1986vz,Buras:2014maa} and there is no space for presenting them here. I just want to make a few 
comments:
\begin{itemize}
\item
Our loop calculations in the meson theory with a cut-off $M=\ord(1\gev)$ include the contributions from 
pseudoscalars and lowest-lying vector mesons and the result can be 
cast in the form of evolution equations. It is remarkable that the structure 
of these evolution equations, in particular the anomalous dimension matrix 
in the meson theory, is very similar to the one in the quark-gluon picture.
This allows to perform an adequate matching between the two evolutions in 
question thereby removing to a large extent  scale and renormalization scheme dependences present 
in the results of Step 1.
\item
The inclusion of vector meson contributions in  \cite{Buras:2014maa} in 
addition to pseudoscalar contributions calculated in  \cite{Bardeen:1986vz} 
is a significant improvement over our 1986 analysis bringing the theory 
closer to data.
\item
The same comment applies to the matching between the quark-gluon and 
meson theory which this time has been performed at NLO in QCD. In this manner 
we could justify equating the physical cut-off $M$ of the truncated meson theory (pseudoscalars ane lowest-lying vector mesons)
with the renormalization scale $\mu$ in the quark-gluon theory.
\end{itemize}

The resulting values 
\be\label{NLO+M}
{\rm Re}A_0\approx (13.3\pm1.0)\times 10^{-8}\gev ,\qquad   {\rm Re}A_2\approx 
(1.1\pm0.1)\times 10^{-8}\gev~. 
\ee
show a very significant improvement over the results in (\ref{NLO}) 
bringing the theory closer to the data in (\ref{N1}) and (\ref{N1a}). In 
particular within the uncertainties of our approach we can claim that 
the experimental value of ${\rm Re}A_2$ has been reproduced. 
The amplitude ${\rm Re}A_0$ has been enhanced in this step by almost a factor of  two relative to the result in  (\ref{NLO}) but it is still by a factor of 
two below the data.  But whereas the calculation of ${\rm Re}A_2$ has been 
completed in this step, in order to complete the calculation of 
${\rm Re}A_0$ we have to include QCD penguin contribution to this amplitude.
This brings us to Step 3.
\subsection{Step 3: QCD Penguins}
As pointed out in \cite{Shifman:1975tn} QCD penguin operators, of which the 
dominant one is
\be\label{penguin}
Q_6=-8\sum_{q=u,d,s}(\bar s_Lq_R)(\bar q_R d_L),
\ee
could play an important role in enhancing the ratio $R$ as in the isospin 
limit they do not contribute to $A_2$ and uniquely enhance the amplitude 
$A_0$. However, in 1975 the relevant  matrix element $\langle Q_6\rangle_0$ 
was unknown within QCD and its Wilson coefficient $z_6$ was poorly known. 
The first large $N$ result for this matrix  element using factorization approach has been obtained in \cite{Buras:1985yx} and  
have been subsequently confirmed in \cite{Bardeen:1986vp,Bardeen:1986uz} by 
using an effective Lagrangian describing the weak and strong interactions of 
mesons in the large $N$ limit. It is given by 
\be\label{eq:Q60}
\langle Q_6(\mu)\rangle_0=-\,4 
\left[ \frac{m_{\rm K}^2}{\ms(\mu) + \md(\mu)}\right]^2
(F_K-F_\pi)\,B_6^{(1/2)} \, , \qquad B_6^{(1/2)}=1,
\ee
where we have introduced the parameter $\bsi$ which  equals unity 
in the large $N$ limit.

While this matrix element is much larger than the matrix elements of 
$Q_{1,2}$, its Wilson coefficient $z_6$ is strongly GIM suppressed at 
scales $\ord(m_c)$  due to the fact that it results from the difference 
of QCD penguin diagrams with charm and up-quark exchanges. If these 
masses are neglected above $\mu=m_c$ then $z_6(m_c)=0$ and its value 
is roughly by an order of magnitude smaller than $z_{1,2}$ at $\mu=0.8\gev$.
In  \cite{Bardeen:1986uz}
an additional (with respect to previous estimates) enhancement of the QCD
penguin contributions to ${\rm Re}A_0$ has been identified. It comes from 
an incomplete GIM cancellation above the charm quark mass. But as the analyses 
in  \cite{Bardeen:1986vz,Buras:2014maa} show, this enhancement is insufficient 
to reproduce fully the experimental value of  ${\rm Re}A_0$. We find 
that the $Q_6$ contribution to  ${\rm Re}A_0$ for $\mu\le 1\gev$ is relevant as it is by a factor of $3$ larger 
than  ${\rm Re}A_2$. Yet at $\mu=0.8\gev$ it contributes only at the 
level of $15\%$ to the experimental value of  ${\rm Re}A_0$.
\subsection{Summary of Results}
Our final results for $K\to\pi\pi$ amplitudes can be summarized as follows
\be\label{NLO+M+P}
{\rm Re}A_0\approx (17.0\pm 1.5)\times 10^{-8}\gev ,\qquad   {\rm Re}A_2\approx (1.1\pm 0.1)\times 10^{-8}\gev, \qquad R\approx 16.0\pm 1.5~.
\ee
Even if the result for ${\rm Re}A_0$ is not satisfactory, it should 
be noted that the QCD dynamics identified by us was able to enhance 
the ratio $R$ by an order of magnitude. We therefore conclude that 
QCD dynamics is dominatly responsible for the $\Delta I=1/2$ rule.

\begin{figure}
\centering
\includegraphics[height=6.0cm]{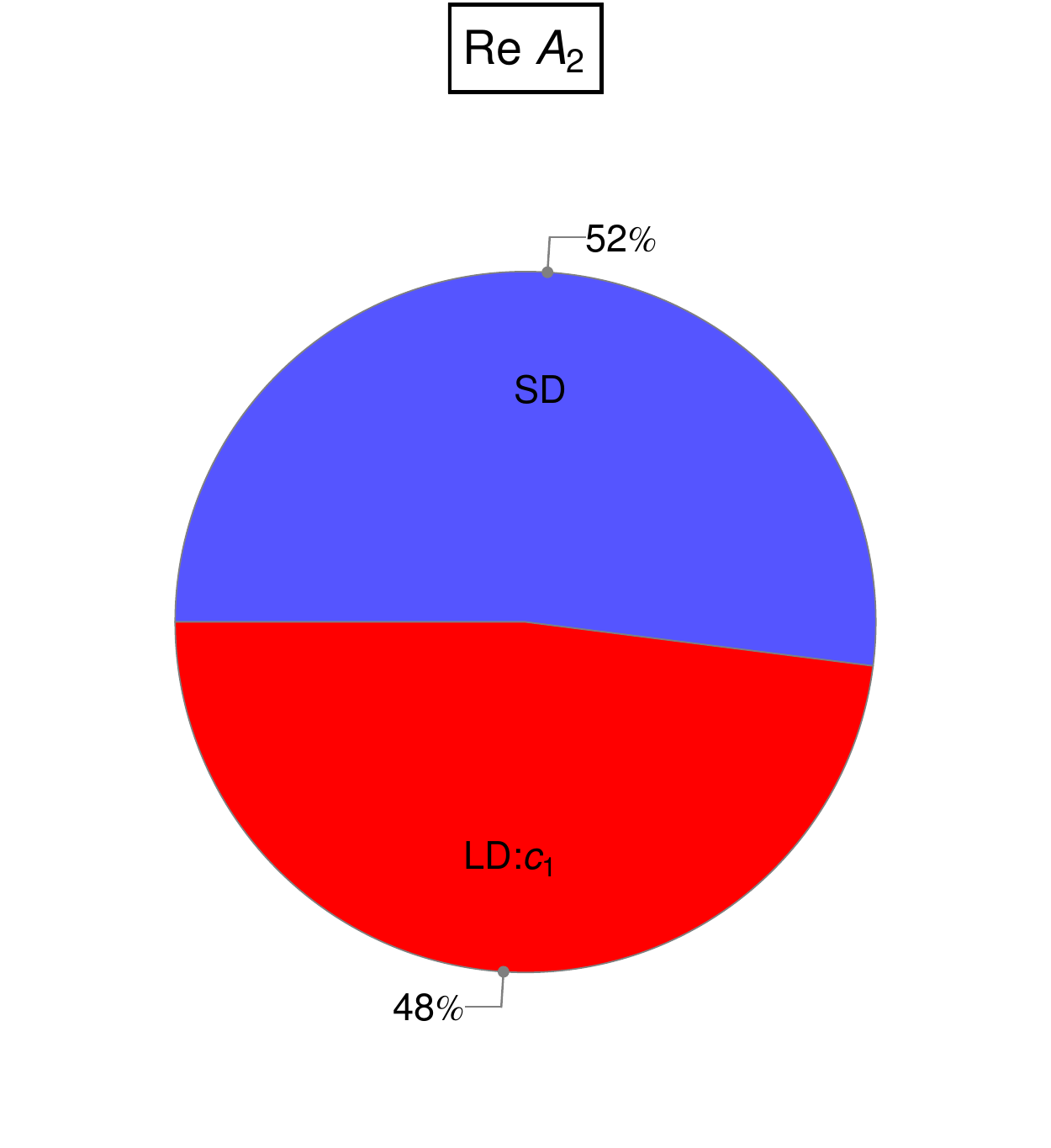}
\includegraphics[height=6.0cm]{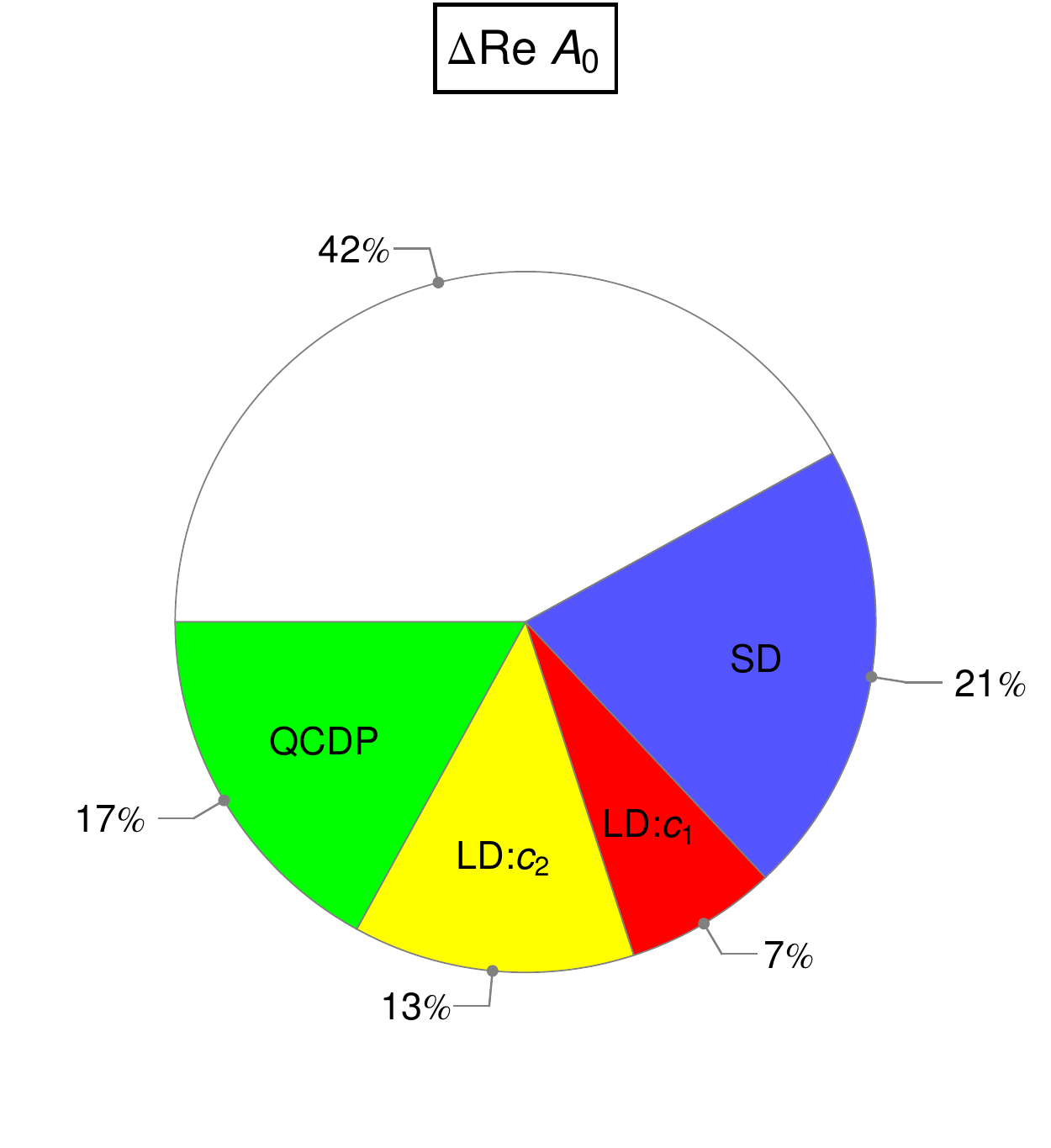}
\caption{Budgets for ${\rm Re}A_2$ (left) and   $\Delta{\rm Re}A_0$ (right) 
summarizing the size of different suppression mechanisms of 
${\rm Re}A_2$ and enhancement mechanisms of ${\rm Re}A_0$, denoted here 
by $\Delta{\rm Re}A_0$, for the matching scale $\mu=M=0.8\gev$. SD stands for 
quark-gluon evolution and LD for meson evolution. In the case of 
 $\Delta{\rm Re}A_0$ we decompose LD into contributions coming from $c_1$ and $c_2$. 
QCDP stands for $Q_6$ contribution. See the text and  \cite{Buras:2014maa} for 
explanations.}
\label{fig:pies}
\end{figure}

In Fig.~\ref{fig:pies} we show budgets for ${\rm Re}A_2$ (left) and ${\rm Re}A_0$ (right) that summarize
the size of different suppression mechanisms of 
${\rm Re}A_2$ and enhancement mechanisms of ${\rm Re}A_0$. SD stands for 
quark-gluon evolution and LD for meson evolution. In the case of 
 ${\rm Re}A_0$ we decompose LD into contributions coming from the meson 
evolution involving only $Q_1$ and $Q_2$ ($c_1$) and the one 
related to the mixing of $Q_{1,2}$ and $Q_6$ ($c_2$). 
QCDP stands for $Q_6$ contribution. We set the matching scale at $\mu=0.8\gev$.
 As can be seen, we are not able to explain fully the missing   $\Delta{\rm Re}A_0=23.4\times 10^{-8}\gev$ relative to the large $N$ limit. Different contributions in the budget are normalized to this additive contribution required by the data. The missing piece that we presently cannot explain by QCD dynamics 
within our approach is represented by the white area. More details on this 
budget can be found in \cite{Buras:2014maa}.

\subsection{Comments on Lattice QCD Results}
Lattice QCD calculations made significant progress in the last five years 
through the inclusion of dynamical fermions \cite{Tarantino:2012mq,Sachrajda:2013fxa}. Among many results the precise 
values for the weak decays constants $F_K$, $F_{B_d}$, $F_{B_s}$ and $F_D$ 
should be mentioned here. The values of the non-perturbative parameters $B_i$
representing  $\Delta F=2$ operators both within the SM and in its extensions 
require further improvements, but it is likely that in this decade they 
will be known with high precision. 

From my point of view, the most important lattice QCD results as far as 
$B_i$ parameters are concerned are the following ones
(recent FLAG update of \cite{Colangelo:2010et} and \cite{Blum:2012uk}):
\be\label{L2012}
\hat B_K=0.766\pm 0.010, \qquad  \bei(3\gev)= 0.65\pm 0.05~, \qquad  ({\rm in~lattice~QCD,}~2014).
\ee
The first one is relevant for the parameter $\varepsilon_K$ and the 
second for the contribution of the dominant electroweak penguin operator
$Q_8$ to the ratio $\epe$. Unfortunately there is no reliable result on $\bsi$ 
in (\ref{eq:Q60}) from lattice QCD so that $\epe$ cannot be calculated in this approach at present.

Concerning $\hat B_K$, the result in (\ref{L2012}) confirmed with 
higher  precision our finding in  \cite{Bardeen:1987vg} that $\hat B_K$ 
is rather close to its large $N$ value $\hat B_K=0.75$. While in 1987, 
including only pseudoscalar meson contributions we found $\hat B_K=0.66\pm 0.07$, our recent calculation that takes into account also vector meson 
contributions and improves the matching between the meson and quark-gluon 
theory gives  \cite{Buras:2014maa}
\be\label{BBGBK}
\hat B_K=0.73\pm 0.02,  \qquad  ({\rm in~dual~QCD,}~2014).
\ee

This result is in an excellent agreement with the lattice QCD value 
in (\ref{L2012}) although we are aware of the fact that while lattice 
calculations have good control over their errors, this is not quite 
the case here. On the other hand, while until now lattice community 
did not provide, as far as I know, any explanation why after 25 years of efforts they 
obtained the result for $\hat B_K$ within $2\%$ from its large $N$ 
value, our approach provides the  explanation why $1/N$ corrections 
are so small. The smallness of these corrections  results from 
 an approximate cancellation between  pseudoscalar  and vector meson one-loop contributions. It is encouraging that such a simple analytic 
approach could provide some insight in the lattice results for  
$\hat B_K$. On the other hand there is a qualitative difference between the results 
in (\ref{L2012}) and (\ref{BBGBK}). While the lattice result finds 
$1/N$ corrections to be {\it positive}, G\'erard has demonstrated diagrammatically in \cite{Gerard:2010jt} that it must be {\it negative}. I expect therefore 
that future lattice results will confirm this result with higher precision 
than we could do it in our approach. 

As far as the $\Delta I=1/2$ rule is concerned, a detailed comparison 
of the results of our approach with the results from  the RBC-UKQCD collaboration \cite{Boyle:2012ys,Blum:2011pu,Blum:2011ng,Blum:2012uk} can be found in 
 section 9 in  \cite{Buras:2014maa}. The results for the amplitudes ${\rm Re}A_0$ and ${\rm Re}A_2$ in lattice QCD 
are presented 
in terms of the  contractions $\circled{1}$ and $\circled{2}$ which are 
depicted in Fig.~1 of \cite{Boyle:2012ys}. Basically, $Q_2$ contributes to $K^0\to\pi^+\pi^-$ and $K^0\to\pi^0\pi^0$ through contractions $\circled{1}$ and $\circled{2}$, respectively, while in the case of $Q_1$ the role of contractions is 
interchanged. The explicit formulae for ${\rm Re}A_0$ and ${\rm Re}A_2$ in terms of these contractions can be found in (122) and (123) in \cite{Buras:2014maa}. 

Now in  \cite{Boyle:2012ys} $\circled{2}\approx -0.7~ \circled{1}$ has been  found.  
This is an important result as it leads to an additional suppression 
of  ${\rm Re}A_2$ and additional enhancement of  ${\rm Re}A_0$ beyond 
the one from quark-gluon evolution, which in  \cite{Boyle:2012ys}
 is stopped at $\mu=2.15\gev$. 
These suppressions and enhancements due to the different signs of 
contractions in question correspond to Step 2 in our approach. Similar 
to the case of $\hat B_K$ the authors of  \cite{Boyle:2012ys} did not 
provide yet the explanation for the relative sign of these two contractions 
while this is possible within our approach. We find  \cite{Buras:2014maa}
\be\label{dic}
\circled{1}= \frac{X_F}{\sqrt{2}}, \qquad \circled{2}= - 0.33 \frac{X_F}{\sqrt{2}}, \qquad X_F=\sqrt{2}F_\pi(m_K^2-m_\pi^2),
\ee
where  the negative sign follows in our approach from 
the proper matching of the anomalous dimension matrices in the meson and 
quark-gluon pictures of QCD. It is also obtained from explicit one-loop calculation in the meson theory
and can also be seen diagrammatically as discussed in \cite{Buras:2014maa}.

Even if with $X_F=0.0298~\gev^3$ the values of the contractions in (\ref{dic}) appear 
at first sight to be much smaller than the ones  presented in \cite{Boyle:2012ys}, it should be noted that lattice groups work with other renormalization schemes and different scales. In fact one can demonstrate, as seen in (\ref{dic}), that in our case the 
factor relating  $\circled{2}$ and $\circled{1}$ must be smaller in 
magnitude.
Therefore the numerical comparison of the results of \cite{Boyle:2012ys}
with ours must also involve the Wilson coefficients $z_i$. 
The fact that our approach and lattice approach predict similar values for 
 ${\rm Re}A_2$ implies the compatibility 
of both approaches as far as $\Delta I=3/2$ transitions are concerned. Indeed 
the lattice result for  ${\rm Re}A_2$ 
in  \cite{Blum:2012uk} reads:
\be\label{A2Lattice}
{\rm Re}A_2= (1.13\pm0.21)\times 10^{-8}~\gev\,,
\ee
where the error is dominated by systematics. This result is 
in agreement with the data and, within uncertainties, with our result. 
We find it remarkable 
that the central value in (\ref{A2Lattice}) differs from our central 
value in (\ref{NLO+M+P})  by only  a few percent. This is still another support 
for the dual picture of QCD. There is no reliable result for ${\rm Re}A_0$ 
from lattice QCD yet but on the basis of present calculations $R\approx 11$, 
still by a factor of two below the data. As QCD penguin contributions 
at $\mu=(2-3)\gev$ are found to be small, we expect that future lattice 
calculations of hadronic matrix elements of $Q_{1,2}$ will imply significantly 
larger values of $R$.

I would like to end this comparison with lattice QCD with a few personal comments:
\begin{itemize}
\item
I find the study of $K\to\pi\pi$ decays in lattice QCD very important 
but as long as lattice calculations of hadronic matrix elements are performed 
at $\mu=(2-3)\gev$ I do not expect that we will gain a satisfactory physical understanding of  the dynamics behind the $\Delta I=1/2$ rule from this approach. 
Obtaining just two numbers for  ${\rm Re}A_0$  and  ${\rm Re}A_2$ from very 
demanding computer simulations without the understanding of the dynamics 
behind them would be rather disappointing after almost 60 years of efforts 
to understand the $\Delta I=1/2$ rule. I believe that combining the physical insight 
on the dynamics behind  the  $\Delta I=1/2$ rule gained through dual QCD 
approach presented above with lattice QCD calculations could eventually 
 completely uncover the puzzles of the 1950s on $K\to\pi\pi$ decays.
\item
On the other hand,  from the present perspective 
only lattice simulations with dynamical fermions 
can provide  precise values of  ${\rm Re}A_{0,2}$ one day, but this may still 
take several years of intensive efforts by the lattice community \cite{Tarantino:2012mq,Sachrajda:2013fxa,Christ:2013lxa}.  Having precise 
SM values for ${\rm Re}A_{0,2}$ would 
determine precisely the room for NP contribution left  not only in  ${\rm Re}A_0$ but also  ${\rm Re}A_2$.  In turn this would give us two observables which could be used to constrain NP. 
\item
While the issue of the  $\Delta I=1/2$ rule is important, in my opinion 
more pressing is the calculation of $\bsi$ as this would allow one to 
constrain a number of NP scenarios with the help of $\epe$.
\end{itemize}

Other applications of large $N$ ideas 
to $K\to\pi\pi$  and $\hat B_K$, but sometimes in a different spirit than our original approach, are reviewed in 
\cite{Cirigliano:2011ny}. I refer in particular to
\cite{Bardeen:1988zw,Bijnens:1990mz,Pich:1995qp,Bijnens:1995br,Bijnens:1998ee,Hambye:1998sma,Hambye:1999ic,Peris:2000sw,Cirigliano:2002jy,Hambye:2003cy,Gerard:2005yk}. A recent review 
of $SU(N)$ gauge theories at large $N$ can be found in \cite{Lucini:2012gg}. 

Finally I hope that the community of lattice experts will {\it eventually} acknowledge the physical relevance of our simple analytical approach and give us credit for a number 
of findings, listed in  \cite{Buras:2014maa}, that they confirmed 28 years 
later. Afterall, our approach provided an insight into the dynamics behind 
the $\Delta I=1/2$ rule and offered the explanation why  $\hat B_K$ is so close to $0.75$. At least three colleagues in Rome \cite{Carrasco:2013jda} gave us credit for the signs of $1/N$ corrections in QCD to $K\to\pi\pi$ matrix elements and $B_K$ that are opposite to the ones obtained using vacuum insertion approximation.

\section{$Z^\prime$, $G^\prime$ Effects in $K\to\pi\pi$}\label{sec:3}
\subsection{$\Delta I=1/2$ Rule}

As we have seen, presently  the value of ${\rm Re}A_0$  within dual QCD approach is by  $30\%$ below the data and even more in the case of lattice QCD. 
While this deficit could be the result of theoretical uncertainties in both approaches, it cannot be excluded that the missing piece in  ${\rm Re}A_0$
comes from NP.  This question has been addressed in \cite{Buras:2014sba} and I will briefly report on the results of this work. 

\begin{figure}[!tb]
\centering
\includegraphics[width=0.4\textwidth] {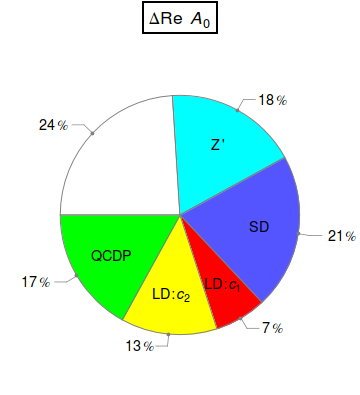}
\includegraphics[width=0.4\textwidth] {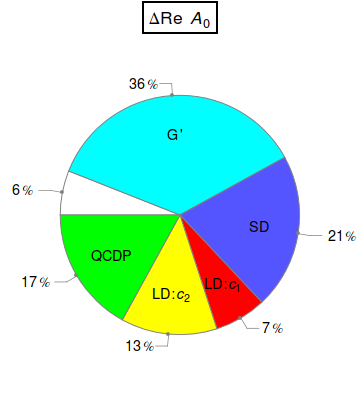}
\caption{\it Budgets of different enhancements of ${\rm Re}A_0$, denoted here 
by $\Delta{\rm Re}A_0$. $Z^\prime$ and $G^\prime$ denote the contributions 
calculated in \cite{Buras:2014sba}. The remaining  coloured 
contributions come from the 
SM dynamics as calculated in \cite{Buras:2014maa} and shown in Fig.~\ref{fig:pies}. The white region stands 
for the missing piece.
}\label{fig:piechart}
\end{figure}

In this paper we have  first demonstrated that a significant part of the 
missing piece in ${\rm Re}A_0$ can be explained by 
tree-level FCNC transitions  mediated by a heavy colourless $Z^\prime$ gauge boson with flavour violating {\it left-handed} coupling  $\Delta^{sd}_L(Z^\prime)$ and approximately 
universal flavour diagonal {\it right-handed} coupling  $\Delta^{qq}_R(Z^\prime)$ to quarks. The approximate flavour universality of the latter coupling assures negligible NP contributions to  ${\rm Re}A_2$.
 This property together with the breakdown of GIM mechanisms at
tree-level allows to enhance significantly the contribution of the leading 
QCD penguin operator $Q_6$ to ${\rm Re}A_0$. A large fraction of the missing piece in the $\Delta I=1/2$ rule can be
explained in this manner for $M_{Z^\prime}$ in the reach of the LHC, while satisfying constraints from $\varepsilon_K$, $\epe$, $\Delta M_K$, LEP-II and the LHC. 
The presence of a small right-handed flavour violating coupling $\Delta^{sd}_R(Z^\prime)\ll\Delta^{sd}_L(Z^\prime)$ and of enhanced matrix elements 
of $\Delta S=2$ left-right operators allows to satisfy simultaneously the constraints from   ${\rm Re}A_0$ and $\Delta M_K$, although this requires some 
fine-tuning. The result of this analysis is summarized by  the left plot in 
Fig.~\ref{fig:piechart}.

 We have also investgated whether a colour octet of heavy neutral gauge bosons 
($G^\prime$) could also help in fully 
explaining the $\Delta I=1/2$ rule. It turns that due to various colour factors 
and different LHC constraints on its mass,  $G^\prime$  is even more effective than 
$Z^\prime$: it provides, within theoretical uncertainties, the missing piece in  ${\rm Re}A_0$ for $M_{G^\prime}=(3.5-4.0)\tev$. Indeed we find 
\be\label{N1aZprime}
R=\frac{{\rm Re}A_0}{{\rm Re}A_2}\approx 18~(Z^\prime),\qquad 
R=\frac{{\rm Re}A_0}{{\rm Re}A_2}\approx 21~(G^\prime)
\ee
with the second result summarized by the right chart in Fig.~\ref{fig:piechart}.

\begin{figure}[ht]
\centering
\includegraphics[scale=0.46]{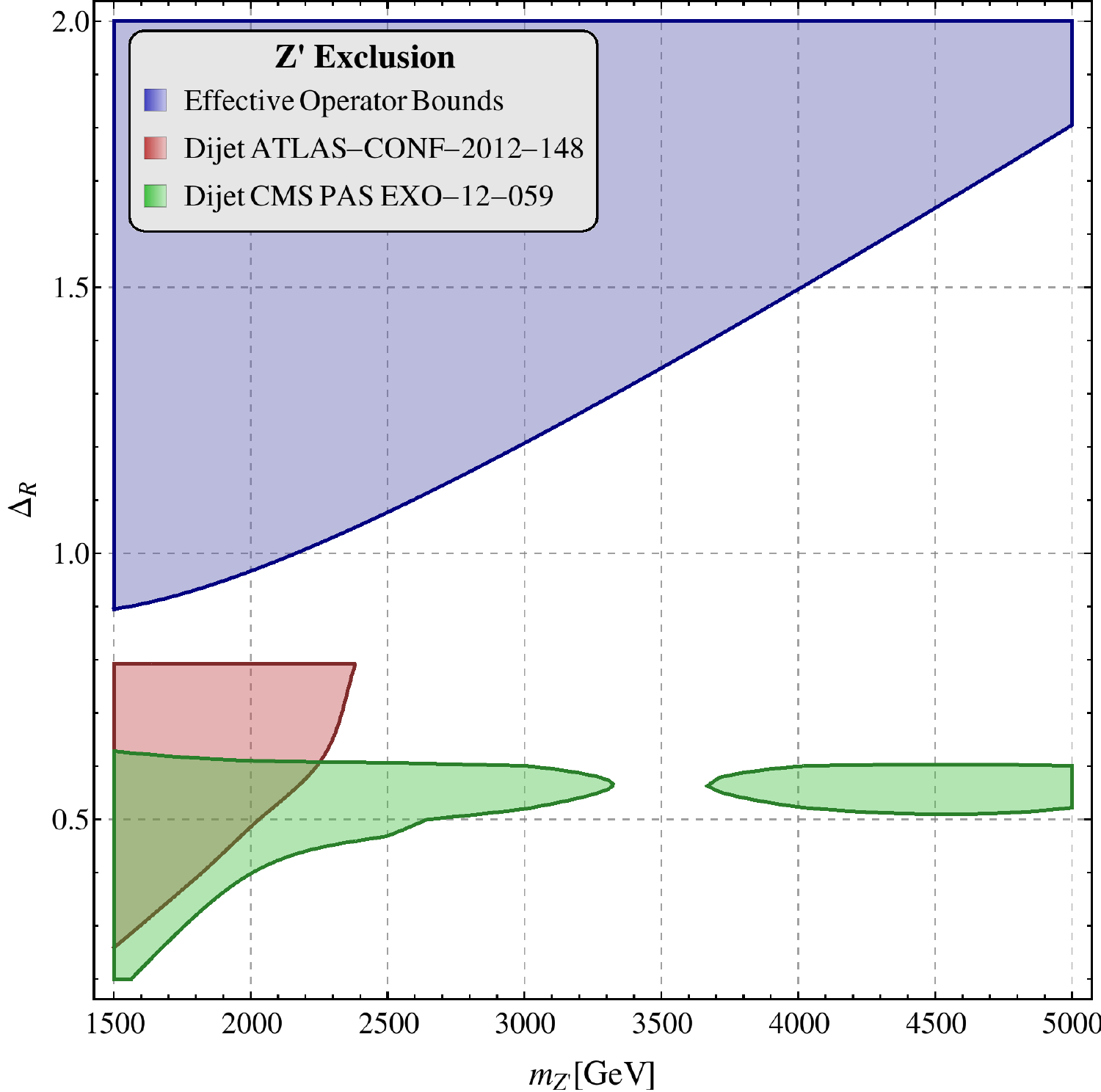}
\includegraphics[scale=0.46]{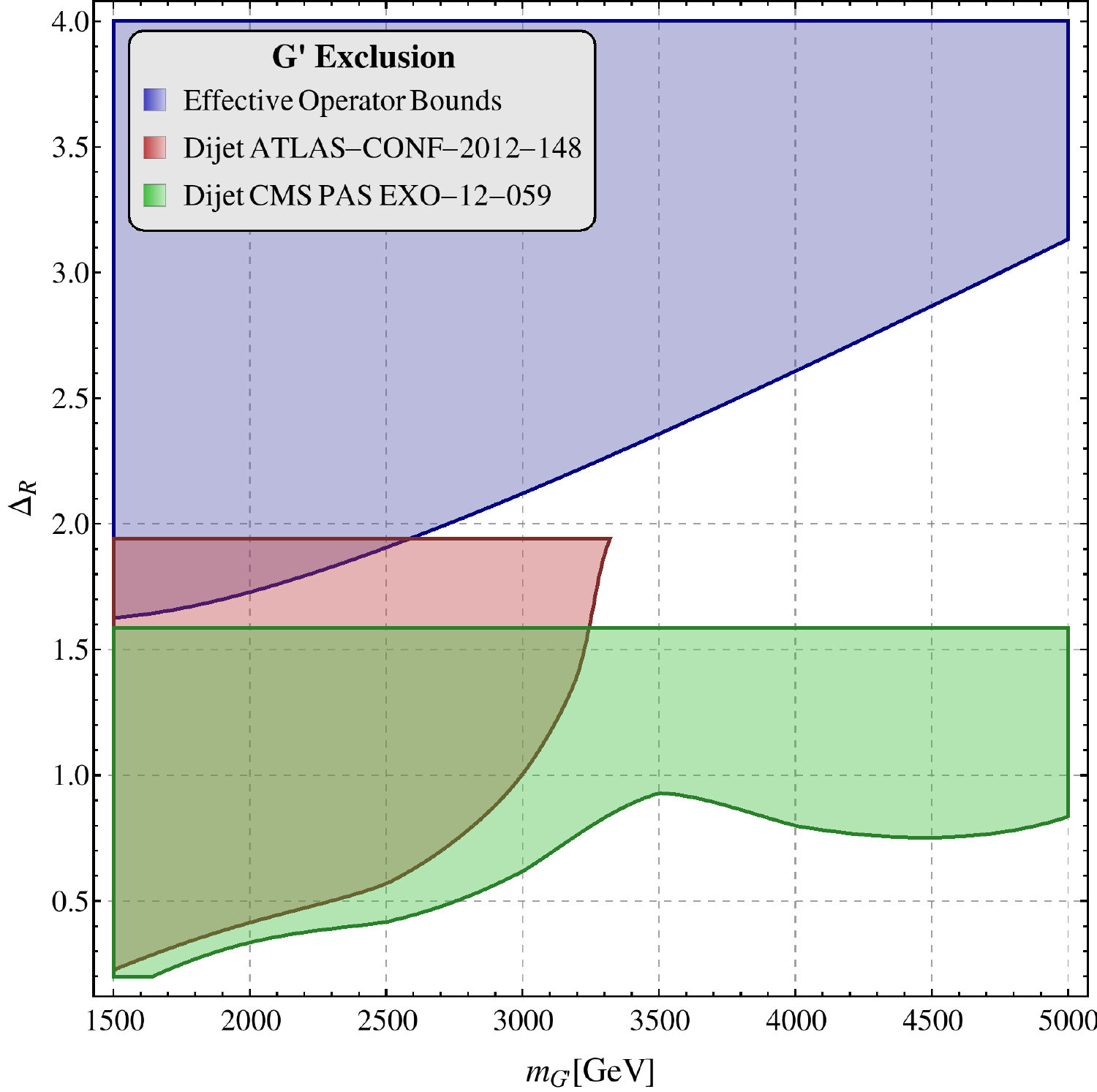}
\caption{\it Exclusion limits for the $Z'$ and $G'$ in the mass-coupling plane, from various searches at the LHC based on \cite{deVries:2014apa}. The blue 
region is excluded by effective operator limits studied by ATLAS \cite{ATLAS:2012pu} and CMS\cite{Chatrchyan:2012bf}. The red and green 
contours are excluded by dijet resonance searches by ATLAS and CMS, respectively.} 
\label{fig:zpgpexclusion}
\end{figure}

The results presented in  \cite{Buras:2014sba} and summarized above used the 
preliminary LHC bounds on the relevant quark couplings provided by 
Maikel de Vries. In \cite{deVries:2014apa} an update on these results has been presented. In particular de Vires points out that the upper bounds on the couplings in the full theory, relevant for our analysis, are slightly softer than the ones following directly from four-quark effective operators at hadron colliders which he provided for our analysis in  \cite{Buras:2014sba}. 
This result not only puts our bounds on the size of NP effects on firm footing but also allows for slightly  larger NP contributions to ${\rm Re}A_0$. Specifically, the bounds  on the relevant couplings in (105), 
(107), (139) and (140) in  \cite{Buras:2014sba} receive additional corrections 
represented by the additional terms between the last square brackets in the formulae below:
\be \left|\Delta_R^{qq}(Z') \right| \leq 1.0 \left[ \frac{M_{Z'}}{3\tev}
 \right] \left[ 1 + \left(\frac{1.3\tev}{M_{Z'}}\right)^2\right]\,,
\ee
\be
\left|\Delta_L^{sd}(Z') \right| \leq 2.3 \left[ \frac{M_{Z'}}{3\tev}
 \right] \left[ 1 + \left(\frac{1.3\tev}{M_{Z'}}\right)^2\right]\,,
\ee
\be
\left|\Delta_R^{qq}(G') \right| \leq 2.0 \left[ \frac{M_{G'}}{3.5\tev}
\right] \left[ 1 +\left(\frac{1.4\tev}{M_{G'}}\right)^2\right]\,,
\ee
\be
 \left|\Delta_L^{sd}(G') \right| \leq 2.6 \left[ \frac{M_{G'}}{3.5\tev} 
\right] \left[ 1 + \left(\frac{1.4\tev}{M_{G'}}\right)^2\right]\,.
\ee
These bounds correspond to the excluded blue regions in Fig.~\ref{fig:zpgpexclusion} and should be compared with the ones in Figs. 3 and 9 in \cite{Buras:2014sba}.

The important feature of these results is that all corrections are above unity. 
In this manner the region representing $Z^\prime$ in Fig.~\ref{fig:piechart} 
can easily be $20\%$ and in the case of $G^\prime$ the white region can be practically removed. Of course all these changes are within the uncertainties of 
the analysis in  \cite{Buras:2014sba} but it is gratifying that the results
 in  \cite{deVries:2014apa} put our analysis on firmer footing. 

Finally, it should be stressed that the allowed ranges for NP contributions in 
 Fig.~\ref{fig:piechart} are independent of $M_{Z'}$ and $M_{G'}$ as with 
increased values of these masses the propagator suppression in  ${\rm Re}A_0$
is compensated by the increase of the allowed ranges for the couplings. The 
additional corrections in the formulae above introduce weak mass dependence for masses above $3\tev$ which should be used in any case to be on the safe side.  Of course one has to stay within the perturbative bounds for the couplings involved. 
This feature tells us that even if $Z'$ and $G'$ would not be found at the LHC, 
they could still play a role in the $\Delta I=1/2$ rule if their masses were 
below $10\tev$. But to find it out would require the study of other observables 
as discussed in  \cite{Buras:2014sba}. Moreover, new bounds from the upgraded 
LHC could further restrict NP contributions to this rule.

\subsection{$\epe$}
In view of the improved value for $\bei$ from \cite{Blum:2012uk} in 
(\ref{L2012}) we have  updated in \cite{Buras:2014sba}
 the value of $\epe$ in the SM stressing various uncertainties, originating in the values of $\vub$ and $\vcb$ and also in the parameter $\bsi$. In 
particular we have found that the best agreement of the SM with the 
data is obtained for $\bsi\approx 1.0$, that is close to the large $N$ limit 
of QCD. In this paper one can also find the impact of $Z^\prime$, $G^\prime$ 
and $Z$ with flavour violating couplings on $\epe$. There is no doubt that in 
the 2020s the ratio $\epe$ could become a star of flavour physics as it was in 
the 1990s.

\section{Conclusions}\label{sec:4} 
I have reviewed the present understanding of the $\Delta I=1/2$ rule that
emerged within the dual approach to QCD as a theory of weakly interacting mesons  for large $N$ already 28 years ago in  \cite{Bardeen:1986vz}
and has been put on a firmer footing recently in  \cite{Buras:2014maa}. 
While lattice QCD will eventually provide much more accurate values for 
${\rm Re}A_0$ and ${\rm Re}A_2$ than it is possible in our approach, our 
 approach provided in my opinion better insight into the dynamics behind 
this rule than it was possible with lattice QCD until now. But the story is not over as we presently do not know whether at a level of $(10-30)\%$  NP could be responsible for the measured value of $R$. Lattice QCD could make an important contribution in answering this question in the coming years.

I am looking forward to improved results on ${\rm Re}A_0$ and ${\rm Re}A_2$ 
from lattice QCD and to possible discoveries of $Z^\prime$ and $G^\prime$ 
at LHC2 in order to see whether these heavy gauge bosons have anything 
to say in the context of the $\Delta I=1/2$ rule. But the most pressing now 
is an accurate evaluation of $\bsi$ by lattice QCD as $\epe$ is much more 
sensitive to NP and very short distance scales than the amplitudes 
${\rm Re}A_0$ and ${\rm Re}A_2$ .

\section*{Acknowledgements}
First of all I thank Bill Bardeen and Jean-Marc G\'erard for a very enjoyable 
collaboration on $\Delta I=1/2$ rule and $\hat B_K$ within the dual QCD approach. It is also 
a pleasure to thank Fulvia De Fazio and Jennifer Girrbach-Noe for a very 
efficient study of $Z^\prime$ and $G^\prime$ effects in $K\to\pi\pi$ decays. 
Special thanks go to Maikel de Vries for providing the LHC bounds in question and numerous very useful E-mails.
Finally I would like to thank the organizers of this workshop for inviting me 
to this interesting event and for an impressive hospitality.
This research was done and financed in the context of the ERC Advanced Grant project ``FLAVOUR''(267104) and carries the number ERC-82. It was also partially
supported by the DFG cluster of excellence ``Origin and Structure of the Universe''.

\bibstyle{woc}
\bibliography{Ballrefs}
\end{document}